\documentclass[mathleft
]{an}
\usepackage{graphicx}
\usepackage{times}
\begin{document}

\Pagespan{789}{}
\Yearpublication{2006}%
\Yearsubmission{2005}%
\Month{11}%
\Volume{999}%
\Issue{88}%

\title{Analysis of the possible Blazhko-effect Cepheid V473 Lyrae}

\author{L. Moln\'ar\inst{1}\fnmsep\thanks{\email{molnar.laszlo@csfk.mta.hu}},
L. Szabados\inst{1}, R. J. Dukes, Jr.\inst{2}, \'A. Gy\H{o}rffy\inst{3}, R. Szab\'o\inst{1}
}
\titlerunning{Analysis on the possible Blazhko-effect Cepheid V473 Lyrae}
\authorrunning{L. Moln\'ar et al.}
\institute{
Konkoly Observatory, Research Centre for Astronomy and Earth Sciences, Hungarian Academy of Sciences, Konkoly Thege Mikl\'os \'ut 15-17, H-1121, Budapest, Hungary
\and
Department of Physics and Astronomy, The College of Charleston, Charleston, SC 29424, USA
\and 
Department of Astronomy, E\"otv\"os University, P\'azm\'any P\'eter s\'et\'any 1/A, H-1117, Budapest, Hungary}
\received{}
\accepted{}
\publonline{later}


\abstract{V473 Lyrae is a peculiar Galactic Cepheid, showing strong amplitude modulation that resembles the Blazhko-effect observed in RR Lyrae stars. We collected data spanning several modulation cycles and started a detailed analysis. The first results indicate that the star shows both amplitude and phase modulations with an average period of 1204 days, but both the cycle length and the strength of the modulation are subjected to considerable variations. A possible quintuplet component in the Fourier spectrum and additional period changes were also detected.  }

\maketitle

\section{Introduction}
The Blazhko-effect, the more or less periodic modulation of the pulsation period and amplitude is a quite common (Jurcsik et al. 2009; Benk\H{o} et al. 2010), although still mysterious phenomenon among RR\,Lyrae stars. It is much less common among other radial pulsators, however. A small fraction of double-mode Cepheids in the Large Magellanic Cloud show definite amplitude and phase modulations (Moskalik \& Ko{\l}aczkowski, 2009). In our Galaxy however, only two candidates are currently known: one is Polaris ($\alpha$~UMi; Bruntt et al. 2008), and the other, more plausible case is V473 Lyrae.

The light variation of V473 Lyr (HR 7308) was discovered by Breger (1969). It soon turned out that the star is a Population I Cepheid with very short period (1.4908 days), pulsating likely in the 2nd overtone (Burki et al. 1986), and it shows quite strong amplitude variations on a time scale between 955-1400 days (Percy \& Evans 1980; Burki et al. 1986); most likely around 1210 days (Breger 1981). Cabanela (1991) derived an average cycle length of 1258 days but concluded that the modulation is not strictly periodic. Finally, based on the \textit{Hipparcos} measurements, Koen (2001) confirmed that the modulation period is about 1200 days, and found only amplitude but no phase modulation. \textit{Hipparcos}, however, covered only 1150 days, barely a single modulation cycle.

Several studies tried to explain the strong modulation of V473 Lyr. Burki \& Mayor (1980) suggested the evolution into or out of the instability strip, while Breger (1981) studied the beating of two close frequencies. Binarity was ruled out by Burki (1984). Instability of the limit cycle (Auvergne 1986), and interactions between modes (Van Hoolst \& Waelkens 1995; Breger 2006) were also proposed, but neither hypothesis was truly able to explain the observations. Last, but not least, the Blazhko-effect, the observed modulation in many RR Lyrae stars was suggested by several authors (Burki \& Mayor 1980; Koen 2001; Stothers 2009). The mechanism behind the Blazhko-effect is still unknown: the shortcomings of the three popular models were reviewed by Kov\'acs (2009). Recently, Buchler \& Koll\'ath (2011) proposed the possibility of radial mode resonances.

\begin{figure*}
\includegraphics[width=170mm]{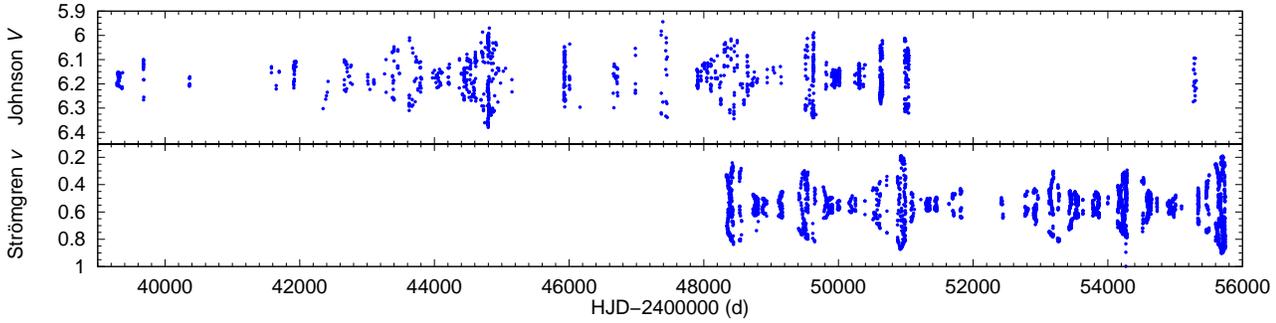}
\caption{Ensemble Johnson $V$ and APT Str\"omgren $v$ light curves, the two best data sets. APT data values are relative magnitudes.}
\label{lc}
\end{figure*}

\begin{figure}
\includegraphics[width=80mm]{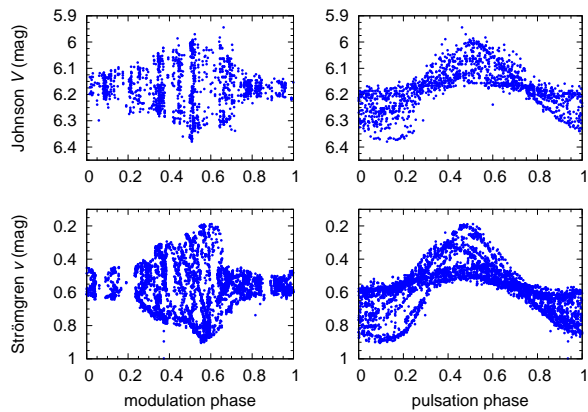}
\caption{Ensemble Johnson $V$ and APT Str\"omgren $v$ light curves, folded with the modulation and pulsation periods: 1204 days and the respective main periodicities from Table \ref{freqs}. Note the high-amplitude portions in the APT data: different modulation cycles show different amplitudes and phase shifts. }
\label{lc_fold}
\end{figure}


\section{Data sets}
We compiled all the available measurements from various sources into a unified data set. Multicolor photometry in Johnson $BVR$ colors and radial velocity measurements were collected. \footnote{References to photometric data sources are: Percy \& Evans (1980); Burki \& Mayor (1980); Breger (1981); Fernie (1982); Burki, Mayor \& Benz (1982); Henriksson (1983); Arellano Ferro (1984); Eggen (1985); Burki et al. (1986); Fabregat, Suso \& Reglero (1990); Arellano Ferro et al. (1990); ESA (1997); Kiss (1998); Ignatova \& Vozyakova (2000); Breger (2006); Berdnikov (2008) and references therein; Szabados (unpublished), Usenko (private comm.); AAVSO data. }

In addition to the publicly available photoelectric and CCD data, we also acquired the complete Str\"omgren $uvby$ observations from the Four College APT (Automatic Photoelectric Telescope) which spans a shorter, partially different time interval but has more even, denser sampling. We first investigated the two longest data sets, the ensemble Johnson $V$ and the APT $v$ data (Figure \ref{lc}). The former is composed of 1573, the latter of 4623 measurements. Because of the color system difference, we did not merge the data sets. 

\section{Pulsation, modulation and variation}
The normal and folded light curves in Figures \ref{lc} and \ref{lc_fold} clearly show the strong modulation of the pulsation. Also, the visual inspection already suggests that the strength of the modulation is not constant. We investigated the strength, period, and in the case of phase variations, the existence of the modulation and also compared it to the Blazhko RR Lyrae stars. 

\begin{table}
\caption{Fourier parameters of the Johnson $V$, APT Str\"omgren and amplitude variation (see Section \ref{ampvar}) data sets. *: $1/f_m$ value.}
\label{freqs}
\begin{tabular}{c c c c c}\hline
ID & Frequency  & Amplitude & phase & $1/ \Delta f$ \\ 
 & (d$^{-1}$) & (mag) & & (d) \\
\hline
\multicolumn{5}{c}{Ensemble Johnson $V$ data}\\
\hline
 $f_1$      & 0.670778 & 0.0743 & 2.028 &  \\
 $2f_1$     & 1.341570 & 0.0107 & 2.535 &  \\
 $f_1-f_m$  & 0.669947 & 0.0322 & 3.246 & 1203.4 \\
 $f_1+f_m$  & 0.671608 & 0.0320 & 0.924 & 1204.8 \\
 $f_2$      & 0.672240 & 0.0202 & 5.103 & 684.0 \\
 \hline
\multicolumn{5}{c}{APT data}\\
\hline
 $f_1$      & 0.670793 & 0.1284 & 3.142 & \\
 $2f_1$     & 1.341426 & 0.0176 & 4.235 & \\
 $3f_1$     & 2.012685 & 0.0049 & 2.713 & \\
 $f_1-f_m$  & 0.669964 & 0.0598 & 3.157 & 1206.3 \\
 $f_1+f_m$  & 0.671618 & 0.0557 & 2.874 & 1212.1 \\
 $2f_1-f_m$ & 1.340603 & 0.0155 & 4.899 & 1215.1 \\
 $2f_1+f_m$ & 1.342236 & 0.0162 & 4.544 & 1234.6 \\
 $f_1-2f_m$ & 0.669124 & 0.0366 & 3.866 & 599.2 \\
$f_1-f_m/2$ & 0.671224 & 0.0382 & 3.359 & 2320.2 \\
\hline
\multicolumn{5}{c}{Amplitude variation data}\\
\hline
$f_m$       & 0.00083080 & 0.0644 & 0.413 & 1203.6* \\
$2f_m$      & 0.00166160 & 0.0232 & 4.675 &  \\
\end{tabular}
\end{table}

\subsection{Frequency analysis} 
\label{ft}
\begin{figure}
\includegraphics[width=80mm]{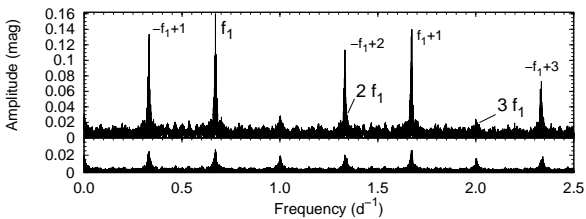}
\caption{Amplitude spectrum of the APT data set. The upper panel shows the full spectrum. Labels indicate the real and alias frequency peak groups. The lower panel shows the residual spectrum after prewithening with the values listed in Table \ref{freqs}.}
\label{sp}
\end{figure}

We performed the Fourier transforms using both the Period04 and TiFrAn packages (Lenz \& Breger 2005; Koll\'ath \& Csubry 2006). Both the pulsation components and the modulation side lobes were identified. However, the  sparse sampling results in strong daily and yearly alias peaks. If the modulation is indeed not constant, it alone creates a forest of sidelobe peaks around the main frequency and its harmonics (Guggenberger et al. 2012) which, coupled with the aliases, makes the frequency determination unambiguous. To make matters worse, the $-f_1+2$ and 2 $\rm{c/d}$ alias groups interfere with the true peaks around the $2f_1$ and $3f_1$ harmonics, as indicated in Figure \ref{sp}.

In the ensemble $V$ data, a single triplet was found at the main ($f_1$) frequency, composed of significant components ($\rm{SNR} > 4$). In the case of the APT data, two triplets around $f_1$ and $2f_1$ and a single quintuplet component at $f_1-2f_m$ were detected. Another peak can be reasonably connected to the $f_1-f_m/2$ value. Although notable residual power is still present in the spectrum, we refrained from overinterpreting it, given the sparseness of the data.

\begin{figure*}
\includegraphics[width=170mm]{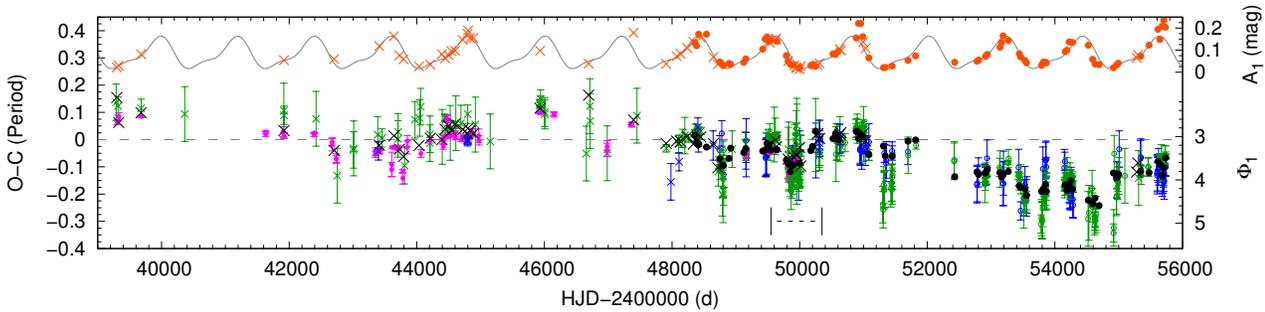}
\caption{$O-C$ diagram, phase and amplitude variations. Orange points: amplitudes of the $f_1$ frequency. The grey line is a two-frequency fit ($f_m$ and $2f_m$). Color codes for the $O-C$ values: green: moments of the median brightness on the ascending branch; blue: maxima; purple: data from Berdnikov \& Pastukhova (1994) and Berdnikov et al. (1997); black: phase variations ($\Phi_1$), inverted to match the $O-C$ values. Crosses: Johnson $V$, dots: scaled APT data ($0.625\,v$). Dashed black line indicates the range of Figure \ref{short}.}
\label{oc}
\end{figure*}

A single additional frequency peak ($f_2$) was identified in the ensemble $V$ data that might be an alias to the $f_1+2f_m$ value but using $f_2$ resulted in a better fit. Modulation periods, based on the positions of the side lobes, are also indicated in Table \ref{freqs}, suggesting an increase from 1204.1 days to about 1213.3 days between the two data sets. In contrast, the pulsation period seem to have decreased, from 1.49081 days to 1.49077 days over a few decades.  

\subsection{Amplitude and phase variations}
\label{ampvar}
\begin{figure}
\includegraphics[width=80mm]{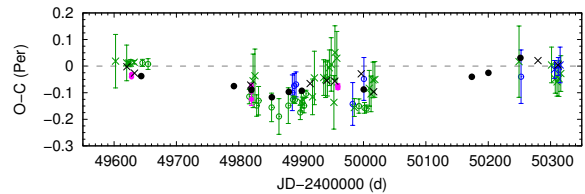}
\caption{$O-C$ diagram during a minimum-amplitude phase, showing signs of short-term phase or light-curve shape variations. Notation is the same as in Figure \ref{oc}. }
\label{short}
\end{figure}

Based on the Fourier transforms, both the pulsation and modulation properties of V473 Lyr seem to be variable. To confirm it, we divided the Johnson $V$ data into 20-80, the APT data into 20-40 day long bins. Each bin was fitted with a single sine function with constant frequency value (0.670775 c/d, from the ephemeris given in Berdnikov \& Pastukhova (1994)), and both the amplitude and phase were adjusted. $O-C$ values were also calculated: we used both the time of maxima and the moments of the median brightness on the ascending branch. Results are summarized in Figures~\ref{oc}, \ref{short} and \ref{loop}. APT amplitudes were scaled down to match the V results.

The calculations revealed that the amplitude modulation shows short, sharp maxima and descending branches, while minima are flatter, with slower ascending branches. The Fourier transform of the amplitude values resulted in a modulation period of $1204 \pm 1$ days, but some cycles deviate from the fitted curve by 50-100 days. The amplitude of the modulation is not constant either: the differing cycles broaden the folded amplitude modulation curve in Figure~\ref{loop}a.

\subsection{Phase variations - changes on different scales}
\begin{figure*}
\includegraphics[width=170mm]{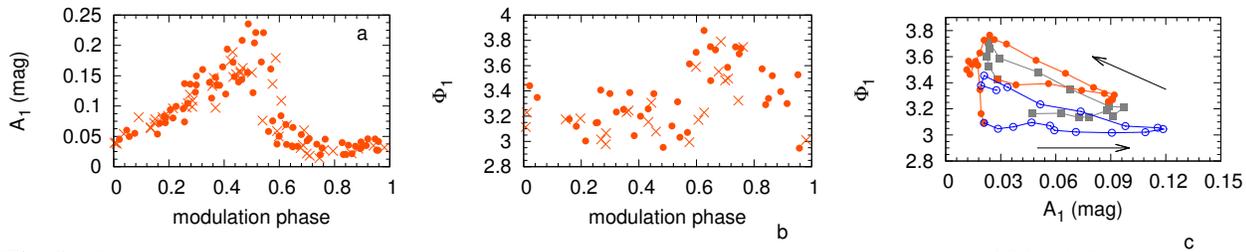}
\caption{Modulation properties: \textit{a.} amplitude variations, folded with the modulation period (1204 d). \textit{b.} phase variations, folded with the modulation period. Only the relatively stable section between 47900-52000 is shown here and in subplot c. \textit{c.} Loop diagram, showing the amplitude and phase relations during the modulation cycles, starting from the grey rectangles, followed by the orange dots and the blue circles. A 4-point moving average was applied to decrease scatter and highlight the counterclockwise rotation. }
\label{loop}
\end{figure*}

We scaled the O-C diagram and the phase variations together by inverting and shifting the latter to the former (Figure~\ref{oc}). The star definitely shows phase modulation in parallel with the amplitude modulation, as shown in Figure \ref{loop}b, contrary to the findings by Koen (2001). But a slower change is also present: the period was definitely decreasing from approximately HJD 2448000 to 2458800, with a rate of $\sim 10^{-8} \rm{d/d}$, but started increasing since then. There may have been a time interval of increased period around HJD 2446000 but the data from that time frame is very sparse. These findings explain the difference found between the average pulsation periods of the two data sets in Section~\ref{ft}. However, the star seem to have exhibited relatively stronger phase variations in the early 20th century, compared to our data sets (Berdnikov et al. 1997).

There are also hints of phase variations in short time scales, when the $O-C$ values and pulsation amplitudes reach the minimum (phase variation reaches the maximum; Figure \ref{short}). The largest effect is in the moments of median brightness on the ascending branch, showing excursions from the phase variation points into both directions. A likely explanation is that the properties of the hump on the ascending branch and/or the pulsation-averaged brightness of the star itself change with the modulation. Varying average brightness (Jurcsik et al. 2008; S\'odor 2009), and hump/bump properties (Guggenberger et al. 2011, 2012) were observed in modulated in RR Lyrae stars. However, the sparse sampling hinders the identification of the hump in the light curve itself.

\subsection{Loop diagram}
The relation and relative contributions of the amplitude and phase modulation can be best visualized on a loop diagram (Figure \ref{loop}c), where we plot $A_1$ vs $\Phi_1$. Because of the scatter of individual data points, a 4-point moving average was applied to highlight the structure and the direction. The modulation cycles show differences, again indicating that the strength of the modulation is not constant. The loop diagram shows similarities that of the (extremely) modulated RR Lyrae star V445 Lyr (Guggenberger et al. 2012).

\section{Conclusions and future work}
We collected and analyzed extensive photometric data of the peculiar Cepheid V473 Lyr, covering several modulation cycles. We identified modulation triplets and a single possible quintuplet component in the Fourier-spectra. We confirmed that the average modulation period is $1204 \pm 1$ days but individual cycles may exhibit deviations both in amplitude and in phase. Contrary to former studies (Koen 2001), we identified phase modulation in parallel to the amplitude modulation of the pulsation. The star shows phase and/or period changes in longer and possibly on shorter time scales. A hydrodynamic survey of second-overtone Cepheids is also planned, similar to the work of Feuchtinger, Buchler \& Koll\'ath (2000), to search for mode resonances and to understand the light-curve shapes. 

In addition to the most extensive data in $V$ and $v$ bands, we acquired $B$, $R$, $uby$ photometry and radial velocity measurements that will allow us to perform more detailed studies of the underlying mechanism and comparisons to Blazhko RR Lyrae stars. Those studies may reveal whether the modulation of V473 Lyr is the same (or very similar) phenomenon as the Blazhko-effect.

\acknowledgements
Fruitful discussions with Zolt\'an Koll\'ath are gratefully acknowledged. This work was supported by the ESTEC Contract No.\ 4000106398/12/NL/KML, the J\'anos Bolyai Research Scholarship and the `Lend\"ulet-2009' Program of the HAS, the European Community's FP7/2007-2013 Programme under grant No. 269194 (IRSES/ASK), the Hungarian OTKA grant K83790 and the HUMAN MB08C 81013 grant of the MAG Zrt. This research was realized in the frames of T\'AMOP 4.2.4. A/1-11-1-2012-0001 ``National Excellence Program -- Elaborating and operating an inland student and researcher personal support system", subsidized by the EU and co-financed by the Europen Social Fund. We acknowledge the AAVSO International Database contributed by observers worldwide.

\end{document}